# On the phase structure of vector-matrix scalar model in four dimensions

V. E. Rochev[a]

State Research Center of the Russian Federation, "Institute for High Energy Physics" of National Research Centre "Kurchatov Institute", Protvino, Russia

**Abstract** The leading-order equations of the $1/N$ – expansion for a vector-matrix model with interaction $g\phi_a^* \phi_b \chi_{ab}$ in four dimensions are investigated. This investigation shows a change of the asymptotic behavior in the deep Euclidean region in a vicinity of a certain critical value of the coupling constant. For small values of the coupling the phion propagator behaves as free. In the strong-coupling region the asymptotic behavior drastically changes – the propagator in the deep Euclidean region tend to some constant limit. The phion propagator in the coordinate space has a characteristic shell structure. At the critical value of coupling that separates the weak and strong coupling regions, the asymptotic behavior of the phion propagator is a medium among the free behavior and the constant-type behavior in strong-coupling region. The equation for a vertex with zero transfer is also investigated. The asymptotic behavior of the solutions shows the finiteness of the charge renormalization constant. In the strong-coupling region, the solution for the vertex has the same shell structure in coordinate space as the phion propagator. An analogy between the phase transition in this model and the re-arrangement of the physical vacuum in the supercritical external field due to the "fall-on-the-center" phenomenon is discussed.

## 1 Introduction

In the present paper we consider a vector-matrix model of the complex scalar field $\phi_a$ (phion) and real scalar mass-less field $\chi_{ab}$ (chion) with interaction $g\phi_a^* \phi_b \chi_{ab}$ in four dimensions $(a, b = 1, \ldots, N)$. This model, known as well as a scalar Yukawa model, is used in nuclear physics as a simplified version of the Yukawa model without spin degrees of freedom, as well as an effective model of the interaction of scalar quarks (squarks). Despite its well-known imperfection associated with its instability (or more precisely, the metastability [1]), this model, as the simplest model of the interaction of fields, often used as a prototype of more substantive theories to elaborate the various non-perturbative approaches in the quantum field theory (see, e.g., [1–8]).

The solution of the equation for the phion propagator in the leading order of the $1/N$ – expansion shows a change of the asymptotic behavior in the deep Euclidean region in a vicinity of a certain critical value of the coupling constant. For small values of the coupling the propagator behaves as free, which is consistent with the wide-spread opinion about the dominance of perturbation theory for this super-renormalizable model. In the strong-coupling region, however, the asymptotic behavior changes dramatically – the propagator in the deep Euclidean region tend to some constant limit. At the critical value of coupling that separates the weak and strong coupling regions, the asymptotic behavior of the propagator $(1/p)$ is a medium among the free behavior and the constant-type behavior in strong-coupling region

The similar change of asymptotic behavior was found also with solution of the system of Schwinger–Dyson equations in some approximations [9,10].

In the present paper we obtain solutions of the nonlinear equation for the phion propagator in the linearized approximation, correctly describing both the asymptotic behavior and the behavior at small momenta, and also investigate the asymptotic behavior of the vertex for zero transfer momentum.

The existence of a critical coupling in the scalar Yukawa model was noticed by practically all authors who have investigated this model using different methods (see, e.g., [3–5] and references therein). This critical constant is generally regarded as a limit on the coupling constant for a self-consistent description of the model by some method. In our approach, however, the self-consistent positive solutions in Euclidean region exist also for the strong coupling, and the existence of the critical coupling looks more like as a phase transition in accordance with the general definition of the phase transition as a sharp change of properties of the model with a smooth change of parameters.

[a] e-mail: rochev@ihep.ru

The structure of the paper is as follows. In Sect. 2, equations of the leading-order of the $1/N$ – expansion for this model and its renormalization are given. In Sect. 3, equation for phion propagator is investigated. It is shown that there exists a critical value of the coupling at which the asymptotic behavior changes. Solutions of the equation in linearized approximation are obtained in three regions: in the weak-coupling region, at the critical value of the coupling, and in the strong-coupling region. In the strong coupling region, the propagator in the coordinate space has a characteristic shell structure. In Sect. 4, an equation for a vertex with zero transfer is investigated. The asymptotic behavior of the solutions in these regions shows the finiteness of the charge renormalization constant. In the strong-coupling region, the solution for the vertex in coordinate space has the same shell structure as the propagator. Discussion and conclusion are contained in Sect. 5. An analogy is made between the phase transition in the model under consideration and the re-arrangement of the physical vacuum in the supercritical external field.

## 2 Preliminaries

The Lagrangian of the model is

$$\mathcal{L} = -\partial_\mu \phi_a^* \partial_\mu \phi_a - m^2 \phi_a^* \phi_a$$
$$- \frac{1}{2}(\partial_\mu \chi_{ab})^2 + \frac{g}{\sqrt{N}} \phi_a^* \phi_b \chi_{ab} \quad (a,b = 1, \ldots, N). \quad (1)$$

The technique of construction of the $1/N$ – expansion for such models is well-known (see [11,12]). In this paper we consider only the leading order of this expansion.

The system of the leading-order equations includes the equations for the phion propagator and the vertex. The leading-order equation for the phion propagator $\Delta_{ab}(x) = \delta_{ab}\Delta(x)$ is

$$\Delta^{-1}(x) = (m^2 - \partial^2)\delta(x) - g^2 D_c(x)\Delta(x). \quad (2)$$

Here $x \in E_4$ and $D_c = -1/\partial^2$.

The phion-antiphion-chion vertex $\Gamma_{ab,cd}(x,y|z)$ has the structure

$$\Gamma_{ab}(x,y|z) \equiv \Gamma_{ab,cc}(x,y|z) = \delta_{ab}\Gamma(x,y|z),$$

where $\Gamma(x,y|z)$ is a solution of the equation

$$\Gamma(x,y|z) = \frac{g}{\sqrt{N}}\delta(x-z)\delta(z-y) + g^2 D_c(x-y)$$
$$\times \int d^4x_1 d^4y_1 \Delta(x-x_1)\Gamma(x_1,y_1|z)\Delta(y_1-y). \quad (3)$$

In the momentum space Eqs. (2) and (3) are

$$\Delta^{-1}(p) = m^2 + p^2 - g^2 \int \frac{d^4q}{(2\pi)^4} D_c(p-q)\Delta(q), \quad (4)$$

and

$$\Gamma(p|k) = \frac{g}{\sqrt{N}} + g^2 \int \frac{d^4q}{(2\pi)^4} D_c(p-q)\Delta(q)\Gamma(q|k)\Delta(q-k). \quad (5)$$

The equation for the chion propagator $D$ is the Dyson equation

$$D = D_c + \frac{g}{\sqrt{N}} D_c * V, \quad (6)$$

(in operator notations), where $V$ is the three-point function. Since $V = O(\frac{1}{\sqrt{N}})$, we see, that $D = D_c + O(1/N)$, i.e., the chion propagator in the leading-order of $1/N$–expansion is the free propagator.[1]

To renormalize the leading order one should to add three counter-terms: $\delta m^2$ (renormalization of the phion mass), $z$ (the phion-field renormalization) and $z_g$ (the coupling renormalization), and in this super-renormalizable model does the only counterterm $\delta m^2$ be infinite

We shall use the normalization at zero momentum

$$\Delta^{-1}(0) = m^2, \quad \left.\frac{d\Delta^{-1}}{dp^2}\right|_{p^2=0} = 1. \quad (7)$$

Here and below $\Delta$ and $m^2$ are the renormalized quantities.

This zero-momentum normalization condition plays a very important role in the construction of the analytic solutions obtained below. When normalizing at another point, the equations become more complicated, and it is hardly possible to solve them without using numerical methods.

The renormalized equation for the phion propagator becomes

$$\Delta^{-1}(p^2) = m^2 + p^2 + \Sigma_r(p^2), \quad (8)$$

where $\Sigma_r(p^2) = \Sigma(p^2) - \Sigma(0) - p^2\Sigma'(0)$ and

$$\Sigma(p^2) = -\bar{g}^2 \int \frac{d^4q}{(2\pi)^4} \frac{\Delta(q)}{(p-q)^2}, \quad (9)$$

Here $\bar{g} \equiv gz_g$, where $g$ is the renormalized coupling.

The vertex also is normalized at zero momenta:

$$\Gamma(0|0) = \frac{g}{\sqrt{N}} = \frac{\bar{g}}{\sqrt{N}} + \bar{g}^2$$
$$\times \int \frac{d^4q}{(2\pi)^4} \frac{1}{q^2} \Delta(q)\Gamma(q|0)\Delta(q), \quad (10)$$

---

[1] In this connection, we note that the leading order of the $1/N$–expansion in this model belongs to the class of so-called quenched approximations. As noted in [1,2] (see also Appendix A of the work [4]), such approximations correspond to a stable vacuum.

and the renormalized equation for the vertex becomes

$$\Gamma(p|k) = \frac{g}{\sqrt{N}} + \bar{g}^2 \int \frac{d^4q}{(2\pi)^4}$$
$$\times \left[ \frac{1}{(p-q)^2} \Delta(q)\Gamma(q|k)\Delta(q-k) \right.$$
$$\left. - \frac{1}{q^2} \Delta(q)\Gamma(q|0)\Delta(q) \right]. \quad (11)$$

## 3 Phion propagator

After the angle integration with the formula

$$\int \frac{d^4q}{(2\pi)^4} \frac{\Phi(q^2)}{(p-q)^2} = \frac{1}{16\pi^2}$$
$$\times \left[ \frac{1}{p^2} \int_0^{p^2} \Phi(q^2) q^2 \, dq^2 + \int_{p^2}^{\infty} \Phi(q^2) \, dq^2 \right] \quad (12)$$

we obtain for the phion propagator the integral equation:

$$\Delta^{-1}(p^2) = m^2 + (1-\lambda)p^2$$
$$+ 2\lambda m^2 \int_0^{p^2} \Delta(q^2) \left(1 - \frac{q^2}{p^2}\right) dq^2, \quad (13)$$

where

$$\lambda \equiv \frac{\bar{g}^2}{32\pi^2 m^2} \quad (14)$$

is dimensionless coupling.

This equation is reduced to the non-linear differential equation

$$\frac{d^2}{(dp^2)^2}\left(p^2 \Delta^{-1}(p^2)\right) = 2(1-\lambda) + 2\lambda m^2 \Delta(p^2). \quad (15)$$

We shall look for the positive solutions ($\Delta^{-1}(p) > 0$) of the equation for the propagator in the euclidean region of momenta. Negative solutions necessarily contain Landau singularities and are therefore physically unacceptable.

In dimensionless variable $t = p^2/m^2$ and for dimensionless function $y(t) = \frac{t}{m^2} \Delta^{-1}$ the Eq. (15) becomes

$$y\ddot{y} - 2(1-\lambda)y = 2\lambda t \quad (16)$$

with initial conditions

$$y(0) = 0, \quad \dot{y}(0) = 1. \quad (17)$$

Depending on the value of $\lambda$, three different types of positive solutions are possible.

*(i) The weak coupling:* $\lambda < 1$. In the weak-coupling region the asymptotic solution at large $p^2$ is

$$\Delta^{-1}(p) = (1-\lambda)p^2 + o(p^2).$$

This asymptotic solution is positive at $\lambda < 1$ and corresponds to the asymptotically-free behavior of propagator.

The approximate analytical solution of Eq. (16) can be found with the linearization procedure:

$$y = y_0 + y_1, \quad y\ddot{y} \approx y_0\ddot{y}_0 + y_0\ddot{y}_1 + \ddot{y}_0 y_1, \quad (18)$$

and the following conditions on the functions $y_0$ and $y_1$ are imposed:

(a) $y_0$ has the right asymptotic behavior and $y_1 = o(y_0)$ at $t \to \infty$;
(b) the initial conditions (17) at $t = 0$ are fulfilled.

In the case of the weak coupling one can choose

$$y_0 = (1-\lambda)t^2 + t, \quad (19)$$

and the equation for $y_1$ will be following:

$$(1 + (1-\lambda)t)\ddot{y}_1 = 2\lambda.$$

This equation can be easily integrated, and the solution of the linearized problem in the weak coupling region is

$$\Delta^{-1}(p^2) = (1-\lambda)p^2 + \frac{1-3\lambda}{1-\lambda}m^2 + \frac{2\lambda}{(1-\lambda)^2}$$
$$\times \frac{m^2}{p^2}\left(m^2 + (1-\lambda)p^2\right) \ln \frac{m^2 + (1-\lambda)p^2}{m^2}. \quad (20)$$

This solution has the right self-consistent asymptotic behavior at $p^2 \to \infty$ and fulfills the normalization conditions (7).

*(ii) The critical coupling:* $\lambda = 1$.

At $\lambda = 1$ the Eq. (16) is the singular Emden–Fowler equation

$$\ddot{y} = 2ty^{-1}. \quad (21)$$

For this equation with above initial conditions the existence and uniqueness theorem has been proved (see [13]).

Equation (21) has exact solution

$$y_{cr} = \sqrt{\frac{8}{3}}\, t^{3/2}, \quad (22)$$

and the corresponding propagator is

$$\Delta_{cr} = \frac{1}{m}\sqrt{\frac{3}{8p^2}}. \quad (23)$$

$\Delta_{cr}$ is the exact solution of homogeneous integral equation, i.e. Eq. (13) at $\lambda = 1$ without inhomogeneous term in the r.h.s. Therefore, $\Delta_{cr}$ is asymptotics of the solution of Eq. (13) at large momenta.

According to this, the asymptotic behavior of propagators in the critical point $\lambda = 1$ has the form

$$\Delta(p) = \sqrt{\frac{3}{8m^2 p^2}}\left(1 + O(1/p^2)\right)$$

at large $p^2$ and drastically differs from the asymptotically-free behavior in the weak-coupling region.

To solve the linearized problem at $\lambda = 1$ one should takes as $y_0$ the function

$$y_0 = \sqrt{\frac{8}{3}} t^{3/2} + t. \tag{24}$$

With such $y_0$ the equation for $y_1$ is

$$t(4t + \sqrt{6t})\ddot{y}_1 + 3y_1 = -3t.$$

The change of variable $x = -\sqrt{8t/3}$ transforms this equation into the hypergeometrical ones, and the solution of the linearized problem in this case is the real-valued positive function

$$y = \sqrt{\frac{8}{3}} t^{3/2} + t F\left(1 + i\sqrt{2}, 1 - i\sqrt{2}; 3; -\sqrt{\frac{8t}{3}}\right). \tag{25}$$

Here $F$ is the Gauss hypergeometrical function [14].

*(iii) The strong coupling:* $\lambda > 1$.

This super-critical case requires some comment. Zero-momentum normalization condition (7) implies that the phion-field renormalization is $z = 1 - \Sigma'(0)$. Using definition (9) and formula (12), it is easy to see that $\Sigma'(0) = \lambda$. Consequently, the phion-field renormalization constant is $z = 1 - \lambda$. In the strong-coupling region $z$ is negative that looks like a pathology.

Let us clarify the problem. For a propagator normalized *on the mass shell*, the negativity of the field renormalization constant $z_m = 1 - \Sigma'(-m^2)$ indicates the existence of problems with unitarity. To solve the problem of the sign of $z_m$, it is necessary to solve the equation for the propagator at an arbitrary normalization point $p^2 = \mu^2$ and then to continue this solution in the point $\mu^2 = -m^2$. The normalization of the renormalized propagator $\Delta(p^2)$ at point $\mu^2$ leads to the renormalized Eq. (8) where the renormalized mass operator $\Sigma_r$ will be

$$\Sigma_r(p^2) = \Sigma(p^2) - \Sigma(\mu^2) - (p^2 - \mu^2)\Sigma'(\mu^2) \tag{26}$$

in the case, and

$$\Sigma'(\mu^2) = \frac{\bar{g}^2}{16\pi^2} \frac{1}{(\mu^2)^2} \int_0^{\mu^2} \Delta(q^2) q^2 dq^2. \tag{27}$$

If we make a formal transition to the mass shell, that is, just put $\mu^2 = -m^2$ in above formula, then we will see that $\Sigma'(-m^2)$ will be negative (provided the $\Delta$ is positive), and, accordingly, $z_m$ will be positive.

This reasoning, of course, is not evidence, but rather a suggestive consideration, because it implies a positivity of the propagator in the pseudo-Euclidean region up to the point $p^2 = -m^2$. The detailed study of the equation for the propagator at an arbitrary normalization point, involving the use of numerical methods, is beyond the scope of this paper. We only note that the positivity of the propagator in the Euclidean region can be regarded as a physical principle, and solutions that violate this principle should be rejected. The change of the sign in the pseudo-Euclidean region in the interval from zero to $m^2$ would mean the singularity of the propagator at some point $m_0^2 < m^2$, which can be interpreted as dynamical generation of a new light particle – a fact in itself is quite interesting. In any case, this point requires further study.

For zero normalization point at $\lambda > 1$ Eq. (16) has the positive exact solution

$$y_s = \frac{\lambda}{\lambda - 1} t, \tag{28}$$

and, correspondingly, Eq. (15) has solution

$$\Delta_s = \frac{\lambda - 1}{\lambda} \frac{1}{m^2}. \tag{29}$$

$\Delta_s$ is the solution of integral equation (13), in which $m^2$ in the inhomogeneous term is replaced by $\frac{\lambda-1}{\lambda} m^2$. Hence, $\Delta_s$ is asymptotics of the solution at $p^2 \to \infty$:

$$\Delta(\infty) = \Delta_s. \tag{30}$$

The linearization of Eq. (16) with $y_0 = \lambda t/(\lambda - 1)$ leads to the equation for $y_1$:

$$t\ddot{y}_1 + a^2 y_1 = 0. \tag{31}$$

Here

$$a = (\lambda - 1)\sqrt{\frac{2}{\lambda}}. \tag{32}$$

The solution of Eq. (31), which fulfills the initial conditions (17), is

$$y_1 = -\frac{\sqrt{t}}{(\lambda - 1)a} J_1(2a\sqrt{t}).$$

Here $J_1$ is the Bessel function.

Correspondingly, for the inverse propagator we obtain

$$\Delta^{-1}(p^2) = \frac{m^2}{\lambda - 1}\left[\lambda - \frac{1}{a}\sqrt{\frac{m^2}{p^2}} J_1\left(2a\sqrt{\frac{p^2}{m^2}}\right)\right]. \tag{33}$$

This function is positive at all $p^2 \geq 0$ and gives the true ultraviolet asymptotics (30) and zero-momentum behavior: $\Delta^{-1}(0) = m^2$.

In above formulae, $m$ is a renormalized mass that does not coincide with the physical mass of the phion $m_\phi$, since we use zero-momentum normalization. To determine the physical mass of a phion, it is necessary go to the pseudo-Euclidean Minkowski space and determine the position of poles of the propagator, i.e. zeros of the inverse propagator. The inverse propagator in Minkowski space for the strong coupling region is[2]

---

[2] The expansion of the propagator (33) into a (convergent) series in powers of $p^2$ contains only integer powers, and the transition to Minkowski space does not present any problems.

$$\Delta_M^{-1}(p^2) = \frac{m^2}{\lambda - 1}\left[\lambda - \frac{1}{a}\sqrt{\frac{m^2}{p^2}} I_1\left(2a\sqrt{\frac{p^2}{m^2}}\right)\right]. \quad (34)$$

Here $p^2 = p_0^2 - \mathbf{p}^2$ and $I_1$ is the modified Bessel function. The phion mass $m_\phi$ is determined from equation $\Delta_M^{-1}(m_\phi^2) = 0$ and, according to the formula (34), is determined from equation

$$\lambda a \frac{m_\phi}{m} = I_1\left(2a\frac{m_\phi}{m}\right). \quad (35)$$

This equation has, for all values of $\lambda > 1$, an unique positive solution $m_\phi$, corresponding to the simple pole of propagator. Thus, for $\lambda = 2$ from of Eq. (35), we obtain $m_\phi = 1.24m$, for $\lambda = 8$ $m_\phi = 0.67m$, and so on.

Strong-coupling propagator (33) has very interesting shell structure in the four-dimensional Euclidean x-space. Using the Mellin–Barnes representation for $J_1$ (see, e.g. [14]) one can calculate the Fourier transform of (33):

$$\Delta^{-1}(x) = \int \frac{d^4p}{(2\pi)^4} e^{-ipx} \Delta^{-1}(p)$$
$$= \frac{\lambda m^2}{\lambda - 1}\left(\delta^4(x) - \frac{m^2}{8\pi^2(\lambda - 1)^2}\delta(x^2 - x_0^2)\right) \quad (36)$$

where the quantity

$$x_0 = (\lambda - 1)\sqrt{\frac{8}{\lambda m^2}}$$

can be considered as the "euclidean radius" of the phion. This radius increases as the coupling increases, which is very natural for the strong-coupling regime.

In the context of the linearized equation considered here it is not clear whether this shell structure an immanent property of the model or is an artifact of linearization. An argument in favor of the first statement is the study of the vertex function carried out in the next section.

In concluding this section, we note that the main effect associated with the non-linearity of the renormalized equation for the propagator is a sharp change in the asymptotic behavior in the deep Euclidean region at the critical point and in the super-critical strong-coupling region. This phenomenon, of course, does not depend on the linearization procedure, which is used to obtain analytic expressions suitable in the whole range of momenta.

## 4 Vertex at zero transfer momentum

In this section, we consider the equation for the vertex function (11) with zero transfer momentum $k = 0$. Our goal will be to prove the finiteness of the charge renormalization constant $z_g$, which is determined from the Eq. (10). It is clear that for this proof, it suffices to prove the convergence of the integral in (10).

After the angle integration with the formula (12) the renormalized equation for the vertex at $k = 0$ becomes

$$\Gamma(p^2) \equiv \Gamma(p|0) = \frac{g}{\sqrt{N}} + 2\lambda m^2$$
$$\times \left\{\frac{1}{p^2}\int_0^{p^2} q^2 dq^2 \Delta^2(q^2)\Gamma(q^2)\right.$$
$$\left. - \int_0^{p^2} dq^2 \Delta^2(q^2)\Gamma(q^2)\right\}. \quad (37)$$

This equation can be reduced to differential equation

$$\frac{d^2}{d(p^2)^2}(p^2\Gamma(p^2)) = -2\lambda m^2 \Delta^2(p^2)\Gamma(p^2) \quad (38)$$

with the initial conditions which follow from the integral equation (37).

The normalization condition (10) after the angle integration becomes

$$\frac{g}{\sqrt{N}} = \frac{\bar{g}}{\sqrt{N}} + 2\lambda m^2 \int_0^\infty dq^2 \Delta^2(q^2)\Gamma(q^2), \quad (39)$$

and the convergence of integral is dependent of the asymptotic behavior of $\Delta$ and $\Gamma$ at $p^2 \to \infty$. With this circumstance we can approximate $\Delta$ in (38) by its ultraviolet asymptotics (see preceding section). The results of this investigation is the following:

*(i) The weak coupling:* $\lambda < 1$. In the weak coupling region approximating the propagator by its asymptotics

$$\Delta^{-1} \approx (1 - \lambda)p^2, \quad (40)$$

we obtain asymptotic solution

$$\Gamma(p^2) \sim const$$

at $p^2 \to \infty$. Such asymptotics of the vertex means the convergence of integral in (39) and, therefore, the finiteness of $z_g$.

*(ii) The critical coupling:* $\lambda = 1$.

In this case the approximation of propagator by its asymptotics (23) leads to asymptotic solution

$$\Gamma(p^2) \sim \frac{1}{\sqrt{p^2}}\sin\left(\frac{\ln(p^2/m^2)}{\sqrt{2}} + \varphi\right),$$

which also ensures the convergence of integral (39) and the finiteness of $z_g$.

*(iii) The strong coupling* $\lambda > 1$. In the strong coupling region the approximation of the propagator by its asymptotics (29) leads to equation for $\Gamma$, which can be reduced to Eq. (31), and the solution of Eq. (37) with propagator $\Delta_s$ has the form

$$\Gamma(p^2) = \frac{gm}{a\sqrt{N p^2}} J_1\left(\frac{2a}{m}\sqrt{p^2}\right). \quad (41)$$

The asymptotic behavior of $\Gamma$ ensures the convergence of integral (39) and the finiteness of $z_g$ also in this case.

In $x$-space we obtain from (41):

$$\Gamma(x) = \int \frac{d^4 p}{(2\pi)^4} e^{-ipx} \Gamma(p^2) = \frac{g}{\sqrt{N} x_0^2} \delta(x^2 - x_0^2), \quad (42)$$

i.e. the same shell structure as for the propagator (see (36)). Note that no linearization was performed in above calculations of $\Gamma$. Therefore, it is reasonable to conclude that this shell structure is not an artifact of linearization, but is related to the ultraviolet behavior of the model.

## 5 Discussion

The equations of the leading order of $1/N$-expansion in the scalar vector-matrix model have self-consistent positive solutions in the Euclidean region not only in the weak-coupling region, (where a dominance of the perturbation theory in this model is obvious), but also in the strong-coupling region.

At $\lambda = 1$ the asymptotic behavior of the propagator $(1/p)$ is a medium among the free behavior $1/p^2$ at $\lambda < 1$ and the constant-type behavior in strong-coupling region $\lambda > 1$. The phion propagator in the strong-coupling region asymptotically approaches to a constant. It is not something unexpected, if we remember the well-known conception of the static ultra-local approximation, or "static ultralocal model" (see [15] and references therein). In this approximation, all the Green functions are combinations of $\delta$-functions in the coordinate space that are constants in momentum space. Of course, this approximation is physically trivial. In contrast to the ultra-local approximation, our solution of linearized approximation has the standard pole behavior for the small momenta.

The essential point of above consideration of the strong coupling region $\lambda > 1$ is the existence of positive exact solution (28) of Eq. (16) in this case. The attention is drawn to the fact that this positive solution also exists when $\lambda < 0$. This range of values of $\lambda$ corresponds to the theory with non-Hermitian interaction in which

$$g \to ig, \quad g^2 \to -g^2.$$

Such models have been widely discussed in recent years both within the framework of quantum mechanics and in quantum field models (see Bender [16] for review and numerous references). These models, although based on non-Hermite interactions, nevertheless, with proper modifications can lead to physically meaningful results. The place and role of this solution in the non-Hermitian version of the model require a separate consideration.

A sharp change of asymptotic behavior in the vicinity of the critical value is a behavior that is characteristic for a phase transition. This phase transition is similar to the phenomenon of re-arrangement of physical vacuum in the strong external field (see [17–19] and refs. therein). As is known the relativistic Coulomb problem for $Z > 137$ ($Ze$ is the nuclear charge) has some of specific features. The Dirac equation with a potential corresponding to a point charge $Ze$ is not correct for $Z > 137$: here the "fall on the center" known from quantum mechanics occurs (see, for example, [20]). This fall on the center is related with the term $1/r^2$ in a potential of the relativistic Coulomb problem and is the main reason for this re-arrangement of the vacuum. The potential $U$, which corresponds to the propagator, is defined by comparing the Born approximation of the non-relativistic quantum theory with the lower approximation of the relativistic theory (see, e.g., Pauli [21], Bethe and Morrison [22], Gross [23]):

$$U(r) \sim -\frac{g^2}{m^2} \Phi(r), \quad (43)$$

where $r = |\mathbf{x}|$, and $\Phi$ is the response of the classical field

$$\Phi(x) = \int dx_1 \Delta_M(x - x_1) J(x_1) \quad (44)$$

on static source

$$J(x) = \delta^3(\mathbf{x}). \quad (45)$$

In Eq. (44) $\Delta_M$ is the propagator in pseudoeuclidean Minkowski space.

For free propagator $\Delta_c = 1/(m^2 - p^2)$ Eqs. (43) and (44) give the Yukawa potential:

$$U(r) \sim -\frac{g^2}{m^2} \frac{e^{-mr}}{r}.$$

At critical value $\lambda = 1$ for propagator (23) we obtain

$$U(r) \sim -\frac{1}{m} \frac{1}{r^2}. \quad (46)$$

It is a potential of "fall on the center". As we see, the asymptotic behavior of the propagator at the critical point $\lambda = 1$ corresponds to the behavior of the potential at small distances which leads to a phase transition in the supercritical field. Despite all the obvious limitation of this analogy, it undoubtedly indicates the related nature of these phase transitions. The phase transition considered in the proposed work can be a quantum field analogue of re-arrangement of the physical vacuum in a strong external field. In this connection, of interest is the further study of this critical phenomenon, and the search for analogs in other models.

**Acknowledgements** Author is grateful to the participants of IHEP Theory Division Seminar for useful discussion and the anonymous referee for their helpful comments.